\documentclass[11pt]{article}

\usepackage[margin=1.1in]{geometry}
\usepackage{booktabs}
\usepackage{graphicx}
\usepackage{xcolor}
\usepackage{fontspec}
\usepackage{listings}
\usepackage[hidelinks]{hyperref}
\usepackage{microtype}
\usepackage{parskip}

\newcommand{\railstate}[1]{\textsc{#1}}

\newfontfamily\railfont{DejaVuSansMono}[
  Extension = .ttf,
  BoldFont = DejaVuSansMono-Bold]
\newcommand{\rc}[2]{\textcolor[HTML]{#1}{#2}}
\newcommand{\rb}[2]{\textcolor[HTML]{#1}{\bfseries #2}}
\definecolor{railbg}{HTML}{07090B}

\title{The Signal Rail: A Deterministic Motion Grammar for\\
Communicating Conversational Agent State in Terminal Interfaces}
\author{Matteo Grella\thanks{The author conceived the Signal Rail, wrote
its specification, defined the research program, reviewed and validated
all results, and bears sole responsibility for the content.
Implementation and manuscript preparation were carried out with
assistance from Claude Fable~5 (Anthropic), operating under
the author's direction: the model implemented the engines and the
conformance harness from the specification, generated the figures, and
assisted in drafting the text.}\\
\normalsize Crisis24\\
\normalsize\texttt{matteogrella@gmail.com}}
\date{August 2026}

\begin{document}
\maketitle

\begin{abstract}
Terminal interfaces to conversational agents report rich internal state (listening, thinking, executing tools, awaiting input, failing) almost entirely through text, while the motion channel beside
it, the one peripheral vision monitors without reading, carries a single
bit: \emph{alive}. We present the \emph{Signal Rail}, a one-row terminal
status instrument that gives that channel a grammar. Four ideas govern
it: spatial semantics (input, processing, and output zones, with
direction as meaning), a motion grammar (one kinetic rule per state,
never color alone), determinism (frames as a pure function of explicit
inputs, golden-frame testable), and honesty (no invented progress or
activity). We contribute a 45-section normative specification, a
reference implementation inside a working full-duplex local voice agent
driven by real signals, and two further engines held byte-identical to
it by a cross-language conformance harness. State distinguishability is
established structurally; behavioral evaluation is outlined as future
work.
\end{abstract}

\section{Introduction}

The status indicator has not kept pace with the systems it reports on.
Command-line and terminal applications increasingly host conversational agents (voice assistants, chat frontends, tool-using autonomous agents) whose interaction loop cycles through states as different from one
another as \emph{recording the user's speech}, \emph{executing a shell
command with measurable progress}, and \emph{having been interrupted
mid-sentence}. The dominant idiom for all of these is the status line, and its
semantic work is done almost entirely by text: today's agent CLIs name
their phases (\textsc{working}, \textsc{thinking}, \textsc{waiting}),
show the running tool and elapsed time, and print errors, all informative provided the user is reading. Beside that text sits a spinner, and across the first-party and third-party tools we surveyed
the \emph{kinetics} stay coarse: motion
marks busy and demands attention, but no tool assigns a distinct kinetic
rule to each state it names.\footnote{Surveyed
August 2026: the OpenAI Codex CLI's terminal UI (label states
\textsc{starting}/\textsc{ready}/\textsc{working}/\textsc{waiting}/\textsc{thinking}
share one braille spinner; \texttt{codex-rs/tui} sources), Google's
Gemini CLI (static title icons; one spinner, frozen while awaiting
confirmation), and the hook-driven monitors Codemux (four states,
pulse/static dot, color-carried) and tmux-agent-indicator (three states,
color-carried, optional decorative sweep). Motion in these tools marks
coarse activity and attention distinctions only.} So everything rides on reading the words: at a glance, in
peripheral vision, or from across the room, listening is indistinguishable from thinking, and both from a stalled network call. The channel that does not
require reading carries one bit: \emph{alive}. When richer displays
are attempted, they tend to
borrow from media players (waveforms, equalizers) or from graphical
progress bars, both of which communicate less than they appear to: a
waveform shows energy, not state, and progress-bar animation can distort
perceived duration~\cite{harrison2007}. More depends on this than it used to: oversight of agentic systems leans on indirect cues and on whatever status the agent shows, and the traces developers
do consult are not always reliable~\cite{dhanorkar2026}; and when commercial
devices do encode richer state vocabularies, as the smart-speaker light rings do, users correctly identified only about a third of the tested
behaviors~\cite{kunchay2021}. Richer state display is needed, and where unlabeled color-and-motion vocabularies have been tested, users could not read them.

This paper describes the \emph{Signal Rail}, a status display for
conversational agents that occupies a single terminal row and aims at a
specific goal: \textbf{after brief exposure, the user should identify the
agent's state from the rail's pattern alone, without reading its label}.
The rail resembles a piece of dedicated electronic
instrumentation: segmented rather than fluid, mechanical rather than organic, restrained rather than constantly animated. It is specified normatively, down to
timing, transitions, degraded rendering modes, and required tests.

The contributions are:

\begin{enumerate}
\item \textbf{A spatial-semantic layout} for agent state: the rail is
divided into input, processing, and output zones (28\,\%/44\,\%/28\,\%)
that mirror the agent's pipeline, and movement direction carries meaning
(Section~\ref{sec:spatial}).
\item \textbf{A motion grammar} assigning each of twelve states a distinct
motion \emph{rule} rather than a distinct color, with an explicit
transition grammar and priority ordering between states
(Sections~\ref{sec:grammar}--\ref{sec:transitions}).
\item \textbf{Determinism as a design principle}: frames are a pure
function of explicit inputs, making UI animation testable with golden
frames and reproducible across runs, terminals, and resizes
(Section~\ref{sec:determinism}).
\item \textbf{Honesty constraints} prohibiting invented progress,
precision, or activity (Section~\ref{sec:honesty}).
\item \textbf{A normative specification and a reference implementation}:
the full 45-section specification (RFC-2119 normative language, exact
snapshot fixtures, required test list) is provided as ancillary material,
and an implementation ships in a working full-duplex local voice agent in
which every implemented state is driven by real agent behavior
(Section~\ref{sec:implementation}).
\end{enumerate}

We frame this as a design and systems contribution in the
tradition of design-pattern and specification papers. The central
usability claim, that states become identifiable from pattern alone, is
motivated by the design (each state differs in zone, geometry, \emph{and}
motion rule) but has not yet been evaluated with users;
Section~\ref{sec:limitations} outlines the study we consider appropriate.

\section{Design principles}
\label{sec:principles}

The rail is governed by eight commitments, both aesthetic and functional, chosen to
evoke dedicated instrumentation from an older machine rather than a
contemporary animated widget:

\begin{itemize}
\item \emph{segmented} rather than fluid,
\item \emph{mechanical} rather than organic,
\item \emph{directional} rather than ambient,
\item \emph{restrained} rather than constantly animated,
\item \emph{functional} rather than decorative,
\item \emph{readable without color},
\item \emph{rectangular} rather than circular,
\item \emph{deterministic} rather than randomly generated.
\end{itemize}

Several familiar displays are explicitly rejected: the rail is not a
spinner, not an equalizer, not a waveform decoration, not a progress-bar
skin, and not an imitation of the KITT scanner. Two of these principles do most of the work.

\textbf{Every state gets a rule, not a color.} Listening \emph{expands};
captured input \emph{collapses}; thinking \emph{reads}; speaking
\emph{emits}; acting \emph{advances}; waiting \emph{freezes}; needs-input
\emph{returns control toward the user}; warning \emph{pulses without
moving}; error \emph{fractures}; interruption \emph{cuts movement off}.
Because the discriminating feature is geometric and kinetic, the display
degrades gracefully: monochrome, 16-color, ASCII-only, and reduced-motion
renderings all preserve the structural distinctness of states
(Section~\ref{sec:accessibility}); whether humans exploit it is the open
question of Section~\ref{sec:limitations}.
The same reasoning leads accessibility guidelines to prohibit color as the sole carrier of information~\cite{wcag21}; the rail applies it uniformly to an animated component.

\textbf{The grammar depends on restraint.} The idle state does not breathe, the
warning state does not scan, and the completion sweep runs exactly once.
An indicator that is always moving cannot use motion to mean anything.
The reasoning comes from calm technology~\cite{weiser1997}: the rail should live at the edge of attention and come forward only when the state \emph{changes}.

\section{Design lineage}
\label{sec:lineage}

The rail's aesthetic is deliberately nostalgic, and the inherited
objects are specific: two from fiction, which supply the look and the
failure modes, and three from working instrument practice, which supply
the discipline.

\textbf{The scanner: charisma without semantics.} The oscillating red
bar enters screen culture as the sweeping eye of the Cylon Centurions in
\emph{Battlestar Galactica} (ABC, 1978)~\cite{bsg1978} and returns four
years later on the nose of KITT in \emph{Knight Rider} (NBC,
1982--1986)~\cite{knightrider}. The kinship is no coincidence: both series were created by
Glen A.\ Larson~\cite{bsg1978,knightrider}, and the design is so
canonical that hobby electronics names the bouncing-LED circuit after
him: the ``Larson scanner''~\cite{larsonscanner}. The problem is that the scanner \emph{is} a heartbeat (constant amplitude, constant period, bidirectional), and a heartbeat conveys exactly one
bit: \emph{alive}. The specification
rejects it by name (``not \ldots{} an imitation of the KITT scanner'')
and corrects the semantics rather than the style: keep what made the object compelling, the segmented red-lit instrument row on a dark ground and the machine presence; then forbid the bounce and make every
property the scanner holds constant carry meaning instead: direction
(flow), amplitude (signal level), position (pipeline stage), stopping
(blockage), fracture (failure). The scanner is a heartbeat; the rail
aims to be an electrocardiogram.

\textbf{The lens: presence without disclosure.} HAL 9000, in
\emph{2001: A Space Odyssey} (1968)~\cite{kubrick1968}, is the opposite
failure rendered with total artistic control: a static red camera lens, built to a brief of ``elegant simplicity'' with no buttons, nothing but the voice and the lens~\cite{epstein2017}, whose dramatic power is its refusal to disclose internal state. HAL is pure presence;
the horror is that the crew has nothing to read. As an agent interface,
the rail is the anti-HAL: continuous, legible state disclosure treated
as an obligation. Mining fiction for interface lessons is an
established design practice~\cite{shedroff2012,schmitz2008}; what
fiction contributes here is not a design to copy but two failure modes to steer between: charisma without semantics, and presence without disclosure.

\textbf{Working instruments: the discipline.} The rail's restraint
principle has aviation doctrine behind it. The 1981 FAA alerting-system
design guidelines, authored jointly by Boeing, Douglas, and Lockheed,
state the objective in capitals: conform to a \emph{quiet, dark} flight
deck when all systems are operating normally, with no alerts unless required for safety or operability~\cite{darkcockpit}; Airbus carries
the same rule as the overhead panel's ``lights out''
philosophy~\cite{airbus1998}. That is the idle rail: nominal is dark,
and illumination is information. From railway signalling the grammar
takes its closed-vocabulary discipline: a signal displays one
\emph{aspect} from a fixed inventory, each aspect carries exactly one
indication, and the display must be readable under adverse conditions
because the stopping distance exceeds the sighting
distance~\cite{pachl}. Twelve states, one rule each, no state
distinguishable only by hue, is the same contract transposed to a
character grid. Finally, the specification's amber and phosphor-green
theme presets are an homage to monochrome CRT terminals, the registered phosphor designations conventionally associated with green (P1) and amber (P3) displays~\cite{phosphor}, and nothing more than that: the nostalgia is confined to the palette, while
behavior is governed by the grammar. The rail inherits its stage
presence from the scanner and its discipline from the annunciator panel
and the signal mast.

\section{Anatomy and spatial semantics}
\label{sec:spatial}

The rail occupies one terminal row with five fields: a fixed-width
uppercase state label, a left cap, the logical rail, a right cap, and an
optional right-aligned auxiliary value (elapsed time, real progress, an
error code):

\begin{lstlisting}
THINKING    [-------->..=..=..=.------]  T+01.8
ACTING      [##############>----------]  052%
\end{lstlisting}

(Inline examples and the specification's fixtures use the ASCII glyph profile; the rail is readable in this paper's plainest figures for the
same reason it is readable on a capability-limited terminal.
Figures~\ref{fig:loop}, \ref{fig:attention}, and~\ref{fig:ladder} show
the display as actually rendered: the Unicode \emph{Instrument Square}
and \emph{Safe Block} profiles under the truecolor palette, every frame
generated by the reference implementation.)

The logical rail is a sequence of fixed-width cells divided into three
conceptual zones mirroring the agent pipeline:

\begin{lstlisting}
INPUT          PROCESSING             OUTPUT / ACTION
<-- 28% -->    <---- 44% ---->        <-- 28% -->
\end{lstlisting}

Zone boundaries are logical and normally invisible; they structure where
activity is \emph{allowed} to appear. Listening activity occurs only in
the input zone; thinking activity primarily in the processing zone;
speech packets originate at the output-zone boundary and travel right.

\textbf{Direction carries meaning.} Normal information flow is
input\,$\rightarrow$\,processing\,$\rightarrow$\,output, so rightward
movement always means forward processing or emission. Leftward implication
is \emph{reserved}: it may appear only when the assistant is returning
control to the user (needs-input), or when an operation is being retracted
(interruption). Movement must never reverse direction for visual variety,
and the thinking state must never perform the continuous left--right
bounce familiar from decorative scanners. The scanner's bounce strips direction of meaning; forbidding it buys the semantics.

An inactive cell renders as a visible track glyph (\texttt{-}), not a
space: a space inside the rail is reserved to mean a deliberate blackout
or broken connection (used by the error state). Even ``nothing is
happening'' is drawn as an intact electrical path, so that \emph{absence}
can carry meaning elsewhere.

\section{The motion grammar}
\label{sec:grammar}

\begin{figure}[t]
\centering
{\setlength{\fboxsep}{10pt}%
\colorbox{railbg}{\parbox{\dimexpr\linewidth-20pt\relax}{%
\railfont\footnotesize
\input{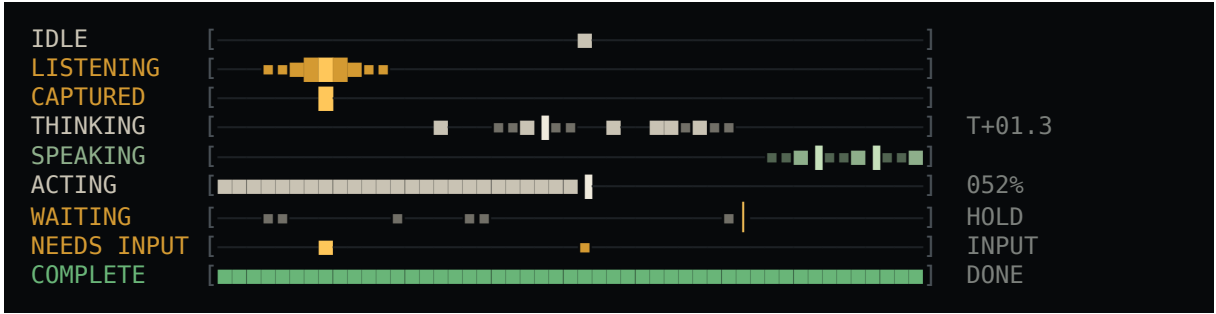}}}}
\caption{The primary conversational loop as rendered: truecolor frames at
the standard 49-cell width on the \texttt{obsidian\_instrument} palette,
generated directly by the reference implementation's \texttt{frame()}
function; each row is one frame at an author-chosen context tuple
recorded in the ancillary generator harness.
\railstate{listening} expands in amber around the input origin;
\railstate{captured} holds its collapsed hot block; \railstate{thinking}'s
read head crosses a seeded sparse field; \railstate{speaking} emits green
packets in the output zone; \railstate{acting} shows determinate progress at an illustrative 0.52
input; \railstate{waiting} freezes dim residue behind an amber
boundary; \railstate{needs input} brightens the input-side marker;
\railstate{complete} has settled its swept rail.}
\label{fig:loop}
\end{figure}

\begin{figure}[t]
\centering
{\setlength{\fboxsep}{10pt}%
\colorbox{railbg}{\parbox{\dimexpr\linewidth-20pt\relax}{%
\railfont\footnotesize
\input{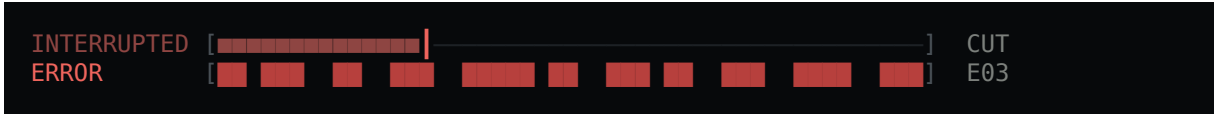}}}}
\caption{Attention states as rendered. \railstate{interrupted} has
retracted to the input boundary and holds its hot cut marker;
\railstate{error} has settled into its seeded fracture: the gaps are
real spaces, an electrically broken rail rather than incomplete
progress. \railstate{warning}, the third attention state, is
specification-only in the voice agent
(Section~\ref{sec:implementation}); the portable engines implement it
from \S25 of the specification.}
\label{fig:attention}
\end{figure}

Twelve states cover the conversational loop: nine \emph{primary} states
tracing the interaction pipeline, and three \emph{attention} states
(\railstate{warning}, \railstate{error}, \railstate{interrupted}) that
can seize the display, per the specification's overlay model
(Section~\ref{sec:transitions}). Table~\ref{tab:grammar} summarizes the
grammar; the fixtures below are the specification's exact 25-cell ASCII
snapshot fixtures, and Figures~\ref{fig:loop} and~\ref{fig:attention}
show the same grammar as a terminal actually draws it.

\begin{table}[t]
\centering
\small
\begin{tabular}{lll}
\toprule
State & Zone & Motion rule \\
\midrule
\railstate{idle}        & processing center & one stable marker; no animation \\
\railstate{listening}   & input             & quantized amplitude expands from an origin \\
\railstate{captured}    & input             & final shape collapses to a compact block \\
\railstate{thinking}    & processing        & read head crosses a seeded sparse field; never bounces \\
\railstate{speaking}    & output            & packets spawn at the boundary, travel right \\
\railstate{acting}      & whole rail        & committed cells + head (real progress), or bounded packet \\
\railstate{waiting}     & frozen            & previous frame freezes; one boundary marker pulses \\
\railstate{needs input} & input+processing  & two markers alternate, handing control left \\
\railstate{complete}    & whole rail        & one rightward sweep, then settled medium rail \\
\railstate{warning}     & whole rail        & fixed lattice, double-pulse; geometry never moves \\
\railstate{error}       & whole rail        & saturate $\rightarrow$ blackout $\rightarrow$ settled fracture \\
\railstate{interrupted} & retracting        & movement stops, retracts left, hard cut marker \\
\bottomrule
\end{tabular}
\caption{The motion grammar: one distinct spatial-kinetic rule per state.
No pair of states differs only by hue.}
\label{tab:grammar}
\end{table}

\begin{lstlisting}
IDLE         [------------=------------]
LISTENING    [.=###=.------------------]
CAPTURED     [--##---------------------]
THINKING     [------->..=..=..=.-------]  T+01.8
SPEAKING     [------------------>#=----]
ACTING       [##############>----------]  052%
WAITING      [##############|----------]  HOLD
NEEDS INPUT  [---=------=--------------]  INPUT
COMPLETE     [=========================]  DONE
WARNING      [#--#--#--#--#--#--#--#--#]
ERROR        [######  ##   ####  ####  ]  E03
INTERRUPTED  [######!------------------]  CUT
\end{lstlisting}

Selected states illustrate how the grammar encodes semantics:

\textbf{Listening vs.\ speaking are structurally different, not mirrored.}
Listening is \emph{localized expansion}: quantized microphone amplitude
$q \in \{0..4\}$ grows a symmetric shape around a fixed origin in the
input zone, with per-cell intensity $\mathrm{clamp}(q{+}1{-}d, 1, 4)$ at
distance $d$. Speaking is \emph{directional packet emission}: short
groups of cells with a bright head spawn at the output boundary and travel
one cell per tick, spawn period governed by the quantized output level (12
ticks at level 0 down to 3 at level 4), at most three packets live, and
overlapping packets keep the maximum intensity per cell. A user can
therefore distinguish ``it hears me'' from ``it is speaking'' by shape
and location alone: the two states never resemble each other even in
monochrome ASCII.

\textbf{Thinking reads; it does not bounce.} The processing zone is
populated with a sparse, structured field of low/medium cells generated
deterministically from $\mathrm{hash}(\mathit{seed}, \mathit{pass},
\mathit{cell})$ with per-cell probabilities targeting the specification's density
ranges (60\,\% track, 25\,\% low, 15\,\% medium in expectation;
individual seeds may fall outside the ranges; no peaks except the head). A read head crosses it left
to right, one cell per two ticks, with a two-cell decaying trail; at the
end of a pass the field dims for one head-step (two base ticks) and a
\emph{new} deterministic field is generated. The metaphor is a head
reading a tape of work. The ``new pass, new field'' rhythm is state
animation, not progress: it marks that thinking persists without claiming
any measurable fraction of work is done.

\textbf{Waiting freezes; interruption cuts; error fractures.} These three
negative-space states are distinct by design. \railstate{waiting}
freezes: a single boundary marker (\texttt{|}) pulses at the point where
progress stopped, and everything else is motionless. (The specification
phrases this as preserving the previous frame; since the frame function
takes no prior-frame input, per Section~\ref{sec:determinism}, implementations
render a deterministic state-local residue field instead, a recorded
deviation that keeps the function pure.) \railstate{interrupted} stops movement immediately, retracts
active cells toward the nearest left semantic boundary over 3--4 ticks,
and leaves a bright hard cut (\texttt{!}) before returning to idle: the
visual of an operation being \emph{withdrawn}. \railstate{error} plays a
one-time entry, full-rail saturation (2 ticks) then blackout (1 tick), and
settles into a fractured pattern whose gaps are real spaces, generated
deterministically from the error code or seed, and which remains until
acknowledged. Failure looks electrically interrupted, not like incomplete
progress; cancellation looks like a cut, not a failure.
Figure~\ref{fig:attention} shows both settled forms.

\section{Transition grammar and priority}
\label{sec:transitions}

Transitions are first-class: one pattern must not be instantly replaced
by an unrelated one. The specification provides a transition table (16
entries) with durations in ticks and visual rules. For example,
\railstate{captured}\,$\rightarrow$\,\railstate{thinking} moves the
compact captured block rightward through exactly three discrete positions
(input origin, zone boundary, processing start), and
\railstate{thinking}\,$\rightarrow$\,\railstate{speaking} places the read
head at the processing/output boundary so that the first speech packet
visibly continues from it: \emph{processed information becomes output}.

Each transition declares interruptibility, and a strict priority order
resolves competing claims:

\begin{lstlisting}
ERROR > INTERRUPTED > WARNING > NEEDS INPUT > WAITING
      > ACTING / THINKING / SPEAKING > COMPLETE > IDLE
\end{lstlisting}

A late-reported failure may replace a completion sweep in flight; nothing
may interrupt an error entry sequence. \railstate{warning} is modeled as
an attention overlay that preserves the underlying state for restoration
after acknowledgement. The specification also permits a primary-state +
attention-overlay state machine, with the constraint that two unrelated
animations never render simultaneously.

\section{Determinism and testability}
\label{sec:determinism}

Every frame is a pure function:

\begin{lstlisting}
frame : (state, entry_tick, tick, width,
         input_q, output_q, progress?, seed, motion) -> cells
\end{lstlisting}

Time is a monotonic tick counter (base rate 12\,Hz, $\approx$83.3\,ms per
tick), never the wall clock. Randomness is prohibited; where variety is
wanted (thinking fields, error fractures), it comes from a hash of a
stable seed (a task or turn identifier) so that a given state at a given
tick renders identically across runs. The logical cell model contains no
escape sequences, no library objects, and no terminal assumptions; a
separate renderer maps cells through a glyph profile and a color mode.

Three engineering properties fall out:

\begin{enumerate}
\item \textbf{Golden-frame testing.} Animations are asserted like any
other output. The specification fixes exact 25-cell ASCII fixtures for
every state (Section~\ref{sec:grammar}) and requires golden tests across
widths, profiles, color modes, and motion modes, with injected elapsed
time and no sleeping in tests. UI animation, usually the least tested code in an application, becomes regression-checked.
\item \textbf{Resize safety.} The specification requires that on resize
the logical state be preserved and the frame \emph{regenerated} at the
new width from normalized positions, never truncated from a rendered
string. (The reference agent sidesteps the hard case by choosing its
width once at startup; live-resize continuity is specified but not yet
exercised by an implementation.)
\item \textbf{Degraded-mode parity.} ASCII and Unicode modes share the
same state machine and differ only in the glyph map, so capability
fallbacks cannot change behavior.
\end{enumerate}

Purity has one recorded cost: rules that reference history (\railstate{waiting}'s frame preservation, \railstate{interrupted}'s retraction of the actual preceding cells) are approximated by
state-local deterministic patterns, because the previous frame is not an
input. An explicit transition-source descriptor is the design
alternative if exact preservation is wanted.

Signal inputs are sanitized and \emph{quantized}: audio-like levels map
into five steps with hysteresis-like limits (rise $\leq$2 levels per
tick, fall $\leq$1, silence settles to zero). The rail responds like an
electronic meter, not a fluid waveform. Quantization both fits the instrument aesthetic and removes the frame-to-frame noise that makes displays
unreadable and untestable.

\section{Honesty constraints}
\label{sec:honesty}

The rail must not display information the system does not have:

\begin{itemize}
\item No percentage is shown unless the backend reports real progress;
unknown progress renders as a bounded work packet with a \texttt{--} or
\texttt{ACTIVE} suffix, never a fake filling bar.
\item Progress never regresses for visual effect; minor backend
fluctuations retain the maximum observed value (an anti-flicker smoothing that can briefly overstate the backend's current claim, a trade the specification accepts and names), and regression is permitted
only on explicit reset, task change, or a labeled new phase.
\item No invented precision in auxiliary values; no artificial activity
when input is silent; no idle animation implying work.
\end{itemize}

Progress displays shape experience: users want
them~\cite{myers1985}, their animation systematically distorts perceived
duration~\cite{harrison2007}, and the distortion can be engineered
deliberately~\cite{harrison2010}. The rail forbids itself these manipulations. For agentic
systems, whose interaction guidelines center on communicating what the
system is doing and how well~\cite{amershi2019}, we advance display honesty as a design value carrying a \emph{hypothesis} that bears on safety, not an established result: an interface that
fabricates progress teaches its user to distrust every other signal it
emits. Establishing the link to calibrated trust requires the studies of
Section~\ref{sec:limitations}.

\section{Accessibility and degraded rendering}
\label{sec:accessibility}
\label{sec:profiles}

The component must remain understandable when colors are unavailable,
motion is disabled, only ASCII is available, or the user cannot
distinguish red from green. Because state identity lives in geometry and
motion, degradation is a matter of \emph{rendering}, not semantics.

\begin{figure}[t]
\centering
{\setlength{\fboxsep}{10pt}%
\colorbox{railbg}{\parbox{\dimexpr\linewidth-20pt\relax}{%
\railfont\footnotesize
\input{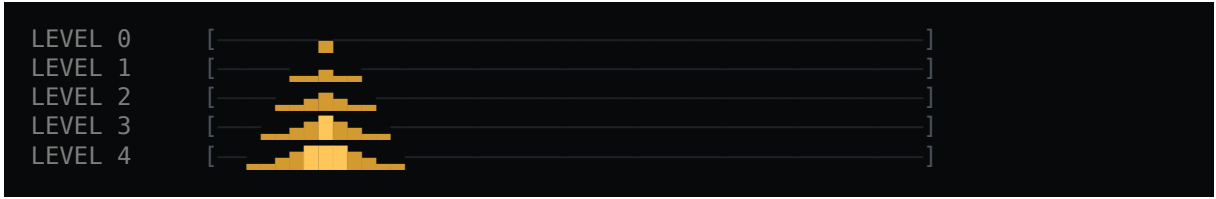}}}}
\caption{The Safe Block profile's listening amplitude ladder, quantized
levels 0--4. Same state machine as Figure~\ref{fig:loop}, different glyph map: profile fallback cannot change behavior, only rendering.}
\label{fig:ladder}
\end{figure}

\textbf{Glyph profiles.} Three profiles share the state machine:
\emph{Instrument Square} (track \texttt{U+2500}, squares
\texttt{U+25AA}/\texttt{U+25A0}, blocks, half-block heads
\texttt{U+2590}/\texttt{U+258C}, boundaries \texttt{U+2502}/\texttt{U+2503});
\emph{Safe Block} (partial-height blocks
\texttt{U+2582}/\texttt{U+2584}/\texttt{U+2586}/\texttt{U+2588}, giving
listening amplitude visible height); and \emph{ASCII}
(\texttt{-\ .\ =\ \#\ >\ <\ |\ !}). Every configured glyph must measure
exactly one terminal column at startup (geometric squares are double-width in some East Asian terminal configurations), with automatic fallback Instrument Square $\rightarrow$ Safe Block $\rightarrow$ ASCII.
Caps are plain ASCII brackets in every profile. Circular glyphs are
prohibited even in ASCII (\texttt{o}, \texttt{O}, \texttt{0},
parentheses). Figure~\ref{fig:ladder} shows why the Safe Block profile exists: partial-height blocks give the listening meter
visible amplitude where geometric squares are unreliable.

\textbf{Color system.} The recommended \texttt{obsidian\_instrument}
theme uses a near-black ground, warm neutrals for processing, amber for
input-side states, restrained green for output, and muted red only for
problems (pure \texttt{\#FF0000} is explicitly avoided). Fallback tables
are specified for xterm-256 and ANSI-16, and a monochrome mode carries
intensity through dim/normal/bold. \texttt{NO\_COLOR}~\cite{nocolor} is
honored. No state may become indistinguishable from another
in monochrome: \railstate{complete} and \railstate{error} differ by
geometry (full rail vs.\ fractures), never only by green vs.\ red.

\textbf{Motion modes.} Three modes are required: \emph{normal};
\emph{reduced} (no continuous travel, transitions in 2--3 discrete
frames, pulses become brightness changes); and \emph{off} (static
patterns that change only on state or data changes, with the text label
authoritative). This mirrors the intent of reduced-motion preferences on
the web~\cite{wcag21} in a medium that has no standard signal for them,
so the specification requires both configuration and a keyboard control.
Flash frequency is bounded: the warning double-pulse produces two
brightenings per 1.4\,s cycle, within WCAG's general
three-flashes-per-second limit~\cite{wcag21}; the criterion's full
threshold conditions (luminance, area) have not been instrumented.

\section{Reference implementation}
\label{sec:implementation}

The rail ships in a full-duplex local voice agent (microphone
$\rightarrow$ streaming STT $\rightarrow$ chat LLM $\rightarrow$ streaming
TTS $\rightarrow$ speaker, with acoustic echo cancellation and barge-in),
implemented in Zig as part of an open-source tensor
framework.\footnote{\texttt{examples/voiceagent/rail.zig} in
\url{https://github.com/matteo-grella/fucina}, pinned at commit
\texttt{9c24016} (${\sim}630$ lines including the golden and
structural test suite); the agent runs entirely on-device on consumer
hardware. The specification, the JavaScript and Python engines, and the
conformance harness accompany the paper as ancillary material and are
maintained at \url{https://github.com/matteo-grella/signal-rail}.} Two properties of the integration are worth reporting.

\textbf{Every implemented state is driven by real behavior.} The agent does not
play-act the grammar. \railstate{listening} amplitude is the quantized
echo-cancelled residual level from the real microphone path, stepped
through the meter-response limiter; \railstate{speaking} packet spawn
follows the RMS of samples actually queued to the audio device;
\railstate{acting} shows determinate progress only for a tool with a real
countdown (a timer), and an indeterminate packet otherwise;
\railstate{waiting} is the agent's pause-then-commit endpointing hold,
frozen behind a pulsing boundary; barge-in triggers the
\railstate{interrupted} retraction and hard cut at the moment playback is
cut; a reply that ends with a question routes
\railstate{complete}\,$\rightarrow$\,\railstate{needs input} until the
user speaks. A fatal error plays the full saturate/blackout/fracture
entry and deliberately leaves the settled fracture above the shell prompt on exit: the failure remains legible after the process is gone.

\textbf{A thin controller implements the transition grammar's timing.} Between the
agent and the pure frame function sits a ${\sim}150$-line controller
that implements the specification's temporal behavior. Transitional
states advance on the clock: \railstate{captured} hands off to
\railstate{thinking} after its four ticks; \railstate{interrupted} and
\railstate{complete} return to \railstate{listening} (or to \railstate{needs input} when the spoken reply asked the user a question), and an unanswered \railstate{needs input} decays to
\railstate{idle} after ${\sim}15$\,s. The
\railstate{idle}$\leftrightarrow$\railstate{listening} boundary is
governed by hysteresis: ${\sim}2$\,s of true silence puts the rail to
sleep, a partial transcript (spoken or typed) pins \railstate{listening}
against premature sleep, and either input energy or the first
transcribed token wakes it within one tick. Level plumbing is thread-safe (the audio producer publishes output RMS through an atomic), and drawing is tick-gated: the pinned row repaints at most once
per 83\,ms tick, plus immediately on state change, satisfying the
specification's performance rules. The rail's width is chosen once at
startup as the largest preset that fits the terminal.

\textbf{The specification survived contact with a real agent, with
recorded deviations.} The implementation deviates deliberately and
documents it: no 256/16-color tiers (truecolor and monochrome only), no
reduced-motion tier (normal, plus static rendering for non-TTY pipes),
glyph-width probing replaced by an explicit ASCII flag, and
\railstate{warning} omitted entirely: this agent has no genuine warning source, and under the honesty stance a state the system cannot truthfully
enter is better absent than simulated. The remaining eleven states, zone discipline, determinism,
quantization bands, and spawn periods are implemented as specified and
pinned by ten deterministic tests: an exact 25-cell golden frame for
every state at pinned context tuples, plus structural properties (zone
partitioning across widths; idle stability; listening containment;
thinking monotonicity and seed-reproducibility; speaking containment;
waiting/interrupted/error monochrome distinctness; acting progress
mapping and non-bouncing; needs-input two-marker phases; quantizer and
meter-response bands). The specification's illustrative 25-cell fixtures
are design targets rather than implementation output; where the
implementation's exact frames differ (the captured block's final width,
for instance) the difference is deliberate and pinned by the goldens.
The per-transition visual bridges of the specification's transition table (the three-position captured hand-off, the thinking-to-speaking head continuation) are \emph{not} implemented: state changes cut on the controller's clock, a recorded deviation and the largest one. The tests run in the project's CI with injected ticks and no
sleeps. Two further engines, HTML/JavaScript for web hosts and dependency-free Python, port the frame function and are
byte-verified against the Zig engine on an 888-frame fixture matrix of
\emph{full logical cells} (glyph, color role, emphasis): $11 \times 4
\times 9 \times 2$ state/width/tick/motion combinations, plus
determinate-\railstate{acting} and 64-bit seed-boundary blocks,
three-way identical and reproducible with one self-contained script
(ancillary material). The seed-boundary block ($0$, $2^{31}{+}1$,
$2^{53}{-}1$, $2^{64}{-}1$) targets the divergences a JavaScript port
invites: bitwise coercion truncates to signed 32 bits, and doubles lose
integer exactness above $2^{53}$. The three-way match is evidence that
the logical-cell contract can hold three languages with three numeric
models to identical output, and that boundary fixtures keep such claims
honest.

The terminal integration pins the rail to the last row via the scroll
region, so conversation text scrolls above an instrument that stays put;
rendering emits compact SGR runs only on style changes.

\section{Related work}
\label{sec:related}

\textbf{Expressive state display on embodied agents.} The closest prior
work is Baraka and Veloso's formalism for revealing a mobile service
robot's internal state through expressive lights~\cite{baraka2018}. The overlap is substantial: they define state
features as predicates over robot variables, cluster them into three
expressible classes (progress toward a known goal, interruption, waiting
for human input), modulate discrete expressions with continuous
variables (percent-done, battery level), resolve co-true features with a
single-winner preference function, and evaluate across design
elicitation, video legibility (a nineteen-point accuracy gain), and a
field experiment in which the lights more than doubled bystander help.
The rail transplants that program to the terminal and sharpens it in
five ways: spatial zones fixed to the conversational pipeline, with
direction reserved for abstract dataflow and control semantics (their
spatial vocabulary is pixel ranges on a strip, with position used ad hoc
for progress fill and turn sides); color-independent degraded modes,
where their expressions rely on color; discrete ticks and stable seeds
in place of wall-clock sampling, so every frame is exactly reproducible;
golden and cross-language conformance tests, which have no counterpart
in their release; and honesty as a stated constraint, where their
real-progress display is an implementation fact rather than a rule.

\textbf{Motion as vocabulary.} That motion itself can carry
iconographic meaning is established. Fairchild et al.\ formalized
icon-to-computational-object mappings, including animated ``automatic
icons,'' in 1989~\cite{fairchild1989}; \emph{Kineticons} built a
39-behavior vocabulary of kinetic manipulations of GUI elements and
validated interpretations with 200 raters~\cite{kineticons}; Harrison
et al.'s point-light study designed 24 temporal behaviors for the most
constrained channel of all, a single LED~\cite{pointlights}. The rail
does not claim this concept; it inherits it. The point-light study's \emph{measured} result, though, is why the rail's medium matters:
eleven target device
states collapsed to only five distinguishable categories, and none of
the 24 tested behaviors rated highly for the ``unable'' family,
suggesting a capacity limit of pure temporal coding. The rail's added channels (position within a semantic zone, glyph shape, a persistent label) are there because of that ceiling.

\textbf{Vehicular light bands.} External HMIs for automated vehicles
put state on one-dimensional light strips, and one deployed concept, Nissan's intention indicator, encoded five vehicle states
partly by sweep direction~\cite{zhang2017}. That precedent is limited in two ways. Its direction is iconic (the light previews where the car will physically move) while the rail's direction is abstract:
rightward is dataflow, leftward is control returning to the user, with
no physical referent. Second, the survey evidence is cautionary: Zhang et
al.'s respondents \emph{reversed} the designers' intended direction
mapping, while in Dey et al.'s light-band study the sweeps were
deliberately symmetric, direction never carried meaning, and animated
patterns did not beat simpler ones~\cite{dey2020}. Direction semantics are learnable, not self-evident, which argues for the rail's persistent labels and for the study of Section~\ref{sec:limitations}.

\textbf{Progress and status indication.} Myers established the value of
percent-done indicators~\cite{myers1985}; Harrison et al.\ showed that
progress-bar behavior systematically distorts perceived
duration~\cite{harrison2007} and can be deliberately engineered to do
so~\cite{harrison2010}. The Signal Rail generalizes the progress bar to a
state \emph{system} and takes a normative position against perceptual
manipulation: display only real progress, and give ``no progress
information'' its own honest form. Practitioner guidance for
command-line tools is adjacent: respond within a beat, show progress for
long operations, decorate with restraint~\cite{clig}.

\textbf{Ambient and calm interfaces.} The restraint principles descend
from calm technology~\cite{weiser1997}: the rail is designed to sit at
the periphery and claim attention only on state change. In Pousman and
Stasko's taxonomy of ambient information systems~\cite{pousman2006} the
rail sits at an unusual point of the design space: peripheral and
low-capacity by design, yet exact rather than abstract in
representation. Unlike ambient
displays, it is an \emph{instrument}: its patterns are specified,
deterministic, and testable.

\textbf{Voice assistant state display.} Commercial voice devices
communicate state through light vocabularies: Amazon's Echo light ring
typically pairs a color with a behavioral pattern per state (directional blue for listening, alternating blue for thinking, pulsing yellow for notification); twelve indicators in the current official
guidance~\cite{alexaring}. These vocabularies are the industrial
ancestor of the rail's grammar, and measured comprehension is poor: across 1{,}006 smart-speaker users, only about 37\,\% of
the tested light behaviors were correctly identified~\cite{kunchay2021}. The rail differs in the things most likely to matter: a persistent text label, geometry and position rather than hue as the primary carrier, and an open specification. It also extends the vocabulary to states that matter for agentic systems~\cite{amershi2019,dhanorkar2026}: tool execution, external waiting, control handback, and interruption.

\textbf{Accessibility guidelines.} The color-independence and motion-mode
requirements apply WCAG's ``use of color'' and animation
guidance~\cite{wcag21} and the \texttt{NO\_COLOR} convention~\cite{nocolor}
for a TUI component, where no browser is present to mediate user
preferences.

\textbf{Specification style.} The normative RFC-2119
vocabulary~\cite{rfc2119} is more familiar from protocol standards than
from UI design documents, though it has been used there before. It separates the binding rules (MUST: distinct rule per
state, no color-only distinction, no fake progress) from taste (SHOULD:
widths, palettes), and it makes the conformance of an independent implementation checkable.

\section{Limitations and future work}
\label{sec:limitations}

\textbf{No user study yet.} The central claim, state identifiability from pattern alone, is a design hypothesis grounded in the grammar's
structure (disjoint zones, disjoint motion rules), not yet a measured
result. The natural evaluation is factorial: display (rail vs.\ spinner) $\times$ label (present vs.\ absent) $\times$ rendering (truecolor, monochrome, ASCII), measuring identification accuracy and latency at
first exposure and after use, with delayed retention. This separates the label's contribution, the geometry's, and their interaction, and distinguishes immediate recognition from learnability. State priors, conversational context,
and expertise need balancing. A complementary field measure is whether
users interrupt or repeat themselves less when the listening/captured
distinction is visible. We
consider this the priority next step, and the deterministic frame
function makes stimulus generation exact and reproducible. The
smart-speaker comprehension study~\cite{kunchay2021} supplies both the
method template and a cautionary baseline: multi-state ambient
vocabularies are routinely misread, and the rail must demonstrate that
structure and labels beat that baseline rather than assume it.

\textbf{Single-row, single-agent scope.} The grammar covers one agent on
one row. Multi-agent orchestration (several rails, or one rail
aggregating subagents), concurrent tool execution, and long-horizon
background work stretch the vocabulary; whether the zone semantics
compose vertically is open.

\textbf{Learned, not self-evident.} The encoding is designed for rapid
learning (each rule is a small physical metaphor: expand, collapse, read,
emit, cut, fracture), but it is still an encoding; first-contact users
must rely on the label. The specification's insistence on labels and on
consistency (``the same visual rule every time a state occurs'') is the
mitigation.

\textbf{Accessibility is visual-only.} The rail's degraded modes cover
every \emph{visual} axis (color, glyph repertoire, motion), but a
screen-reader user gets nothing usable from it: a repainting row of
box-drawing and block glyphs is either skipped or narrated as glyph
names, and every channel the rail uses (position, direction, shape) is
invisible to assistive technology. The needed companion is a non-visual
status channel that announces state \emph{changes} once, as text, per transition. No such channel is yet specified, in this specification or any other we know of for terminal status displays. Until it exists and is tested
with screen-reader users, the rail's accessibility claims must be read
as visual-accessibility claims.

\textbf{Unbuilt transition bridges.} The per-transition visual bridges are the largest specification-to-implementation deviation (Section~\ref{sec:implementation}): they would make state \emph{causality} visible, as when the captured block travels into the processing zone. They are future implementation work, and their legibility value is untested.

\textbf{Terminal-medium limits.} Cell-width probing, font coverage, and
terminal quirks are handled by fallback, but a terminal cannot read user
OS-level reduced-motion preferences; the rail must be told.

\section{Conclusion}

The Signal Rail treats the status row of a conversational agent as an
instrument: a one-row display in which space and motion carry the
semantics of the agent loop. Each state has a rule, not a color:
listening expands, captured input collapses, thinking reads, speaking
emits, acting advances, waiting freezes, errors fracture, interruption cuts. The display therefore survives monochrome, ASCII-only, and
reduced-motion rendering. Every frame is a pure function of state,
entry time, elapsed ticks, width, quantized signal levels, and a seed. Purity makes animation testable like any other program output, and held three independent implementations, in three languages with three numeric models, to byte-identical frames. The rail also never invents progress, precision, or activity that the underlying system does not report. The normative specification (45 sections, exact
fixtures, a required test list), the reference implementation embedded
in a working full-duplex voice agent (listening amplitude is the real echo-cancelled microphone level, progress is a real countdown, barge-in cuts the rail mid-reply), and the conformance harness are
available as ancillary material and open source
(\url{https://github.com/matteo-grella/signal-rail}). We offer the grammar (zones with directional meaning, one motion rule per state, determinism for testability, honesty about progress) as a foundation for terminal-native agent interfaces, and the
identifiability hypothesis as a concrete, falsifiable target for the
evaluation this paper does not yet provide.

\end{document}